\documentclass[letter]{ptptex}

\usepackage{graphicx}

\title{
Universal scaling for the jamming transition%
}

\author{
Michio  Otsuki$^{1,2,}$\footnote{E-mail : otsuki@phys.aoyama.ac.jp} 
and Hisao Hayakawa$^{1,}$\footnote{ E-mail: hisao@yukawa.kyoto-u.ac.jp}
}

\inst{
$^1$ Yukawa Institute for Theoretical Physics, Kyoto University,  
Kitashirakawaoiwake-cho, Sakyo-ku, Kyoto 606-8502, JAPAN
\\
$^2$ Department of Physics and Mathematics, Aoyama Gakuin University,
5-10-1 Fuchinobe, Sagamihara, Kanagawa, 229-8558, JAPAN\footnote{Present adress} 
}

\abst{

The existence of universal scaling in the vicinity of the
jamming transition of sheared granular materials is predicted
by a phenomenology.
The critical exponents are explicitly determined, which are independent of
the spatial dimension. The validity of the theory is verified by the molecular dynamics simulation.

}

\begin{document}

\maketitle

Jamming is an athermal phase transition between the solid-like jammed phase and the liquid-like unjammed phase of granular assemblies.
Above the critical density, which is referred to the point J, the assemblies obtain
 the rigidity and the dynamic yield stress,
while assemblies behave like  dense liquids below the point J.
Liu and Nagel \cite{Liu} indicated that the jamming transition is a key concept of glassy materials. 
Since then many aspects of similarities between the conventional glass transition and the jamming transition have been
investigated.\cite{miguel} \
Indeed, there are many examples where granular materials are used in order 
to investigate dynamical heterogeneity 
in glassy materials.\cite{Daushot05,Abate07,Daushot08,Watanabe} \
On the other hand, we still do not have an unified view in describing glassy materials because
we cannot use the conventional theoretical 
tool for the glass transition such as the mode-coupling theory.\cite{Hayakawa}
 
The jamming is a continuous transition that bulk and shear moduli become nonzero, and there are scaling laws
in the vicinity of the point J similar to the cases of conventional critical phenomena.\cite{OHern02,OHern03} \
Olsson and Teitel \cite{Olsson} and Hatano \cite{Hatano08_1} further demonstrated the existence of 
beautiful scalings near the point J. Therefore we can expect the existence of 
a simple theory in describing the jamming transition.
However, 
we still do not have any theory to determine the critical exponents
of the jamming transition.

In this letter, we predict the critical exponents for jamming transition of sheared granular materials
based on a phenomenology.
First, we introduce the system we consider and
the scaling laws, some of which were introduced in Ref.~\citen{Hatano08_1}. \
Second, based on a phenomenological theory,
we decide the critical exponents for the jamming transition.
Finally, 
we verify the theoretical prediction from our simulation.
The most surprising in our finding is that the critical exponents are independent of the spatial dimension $D$,
 but depend on
the type of particle interaction.

Let us consider a dense sheared and frictionless granular system in which uniform shear flow is stable.
The system  
consists of $N$  spherical grains in $D$ dimensions.
An important parameter to characterize the system is the volume fraction $\phi$.
In contrast to granular gases, the contact force plays crucially important roles in jamming transition, 
where the normal contact force is repulsive one characterized by  $k\delta_{12}^\Delta$. Here, $k$ is the stiffness constant, and
 $\delta_{12}=r - (\sigma_1 + \sigma_2)/2$ with the relative distance $r$ between the
contacting particles of the diameters $\sigma_1$ and $\sigma_2$.
We believe that Hertzian contact law $\Delta=3/2$ is appropriate for three dimensional grains, but we often use
a simpler linear spring model with $\Delta=1$. In this letter, we omit any tangential contact force between grains.
Thus, granular particles are frictionless, which can simplify the argument.

As introduced in Ref.~\citen{Hatano08_1}, 
this system is expected to exhibit the scalings 
for the granular temperature $T$ and the shear stress $S$ in the sheared plane
near the jamming transition as  
\begin{eqnarray}
T & = & A_{T,D}|\Phi|^{x_{\Phi}} {\cal T}_{\pm}\left(t_D\frac{\dot\gamma}{|\Phi|^{x_{\Phi}/x_{\gamma}}}\right), \label{T:scale} \\
S & = & A_{S,D}|\Phi|^{y_{\Phi}}{\cal S}_{\pm}\left(s_D\frac{\dot\gamma}{|\Phi|^{y_{\Phi}/y_{\gamma}}}\right), 
\label{S:scale}
\end{eqnarray}
where $\dot\gamma$ is the shear rate,
$\Phi\equiv \phi-\phi_J$ is the excess volume fraction from the critical fraction $\phi_J$ at the point J, 
$A_{T,D}$ and $A_{S,D}$ are respectively amplitudes of the granular temperature and the shear stress.
The scaling functions ${\cal T}_+(x)$ and ${\cal S}_+(x)$ above the point J respectively differ from ${\cal T}_-(x)$ and ${\cal S}_-(x)$
below the point J.  
It should be noted that $A_{T,D}$, $A_{S,D}$, $t_D$ and $s_D$ do not depend
on  $\dot{\gamma}$ and  $\Phi$,
but depend only on  $D$.

We also introduce a characteristic time scale of sheared granular assemblies as
\begin{eqnarray}\label{omega}
\omega \equiv \frac{\dot{\gamma}S}{nT},
\end{eqnarray}
where $n$ is the number density of grains.
This $\omega$ is reduced to the collision frequency in the unjammed phase 
in a steady state achieved
by the balance between
the viscous heating and the collisional energy loss. 
In contrast to the assumption by Hatano {\it et al.} \cite{Hatano07}, 
$\omega$ also satisfies the scaling form 
\begin{equation}
\omega = A_{w,D}|\Phi|^{z_{\Phi}}{\cal W}_{\pm}\left(w_D \frac{\dot\gamma}{|\Phi|^{z_{\Phi}/z_{\gamma}}}\right) .  \label{w:scale} 
\end{equation}
Thus, there are six critical exponents $x_\Phi$, $x_\gamma$,
$y_\Phi$, $y_\gamma$, $z_\Phi$, and $z_\gamma$
in eqs. \eqref{T:scale}, \eqref{S:scale} and \eqref{w:scale}. 
We note that the normal stress $P$ also satisfies a similar scaling relation
$P=A_{p,D}|\Phi|^{y'_\Phi}{\cal P}_{\pm}(p_D\dot\gamma/|\Phi|^{{y'}_{\Phi}/{y'}_{\gamma}})$
\cite{Hatano08_1},
 but we omit the details of the arguments on ${y'}_{\Phi}$ and ${y'}_{\gamma}$
in this letter. We will discuss them  elsewhere.

Below the point J, 
Bagnold's scaling should be held.\cite{Bagnold,Pouliquen,namiko}  \
Thus, the scaling functions in the unjammed branch satisfy 
\begin{equation}\label{bagnold}
\lim_{x\rightarrow 0}{\cal T}_{-}(x) = 
\lim_{x\rightarrow 0}{\cal S}_{-}(x) = x^2, {\quad}
\lim_{x\rightarrow 0}{\cal W}_{-}(x) = x.
\end{equation}
On the other hand, the jammed branch is characterized by the dynamic yield stress and the freezing of motion.
Then, the scaling functions in the jammed branch satisfy  
\begin{equation}\label{jammed}
\lim_{x\rightarrow 0}{\cal S}_{+}(x) =
\lim_{x\rightarrow 0}{\cal W}_{+}(x) = 1, 
\quad \lim_{x\rightarrow 0}{\cal T}_+(x)=x.
\end{equation}
To obtain the last equation in eq. \eqref{jammed}, we have used eq. \eqref{omega} and the other two equations in eq.
\eqref{jammed}.
Since the scaling functions are independent of $\Phi$  at the point J,  we obtain 
\begin{equation}\label{critical}
\lim_{x\rightarrow \infty}{\cal T}_{\pm}(x) \propto x^{x_\gamma}, \
\lim_{x\rightarrow \infty}{\cal S}_{\pm}(x) \propto x^{y_\gamma}, \
\lim_{x\rightarrow \infty}{\cal W}_{\pm}(x) \propto x^{z_\gamma}.
\end{equation}

Now, let us determine the six critical exponents.
First, we note that there are three trivial relations among the exponents. 
From eqs. (\ref{T:scale})-(\ref{w:scale}) and (\ref{jammed}) we obtain
\begin{equation}\label{z_phi}
z_{\Phi}=y_{\Phi}-x_{\Phi}(1-x_{\gamma}^{-1}).
\end{equation}
Similarly, from eqs.\eqref{T:scale}-\eqref{w:scale} with eq. (\ref{bagnold}) or \eqref{critical} we respectively obtain
\begin{eqnarray}\label{zyx}
z_{\Phi}(1-z_{\gamma}^{-1})&=&y_{\Phi}(1-2y_{\gamma}^{-1})-x_{\Phi}(1-2x_{\gamma}^{-1}), \\
z_{\gamma}&=&y_{\gamma}-x_{\gamma}+1. \label{z_g}
\end{eqnarray}
Thus, we further need three relations to determine the exponents.

In order to introduce the other relations, we consider the 
pressure $P$ in the limit $\dot{\gamma} \rightarrow 0$. 
Let us consider Cauchy's stress in the jammed phase in which 
the pressure $P$ is given by
$P=\sum_{i>j,j} \langle F_{ij}r_{ij}\rangle/V$, where $V$ is the volume of the system,
 $r_{ij}$ and $F_{ij}$ are respectively the distance and the force 
 between $i$ and $j$ particles. This expression may be approximated by
$P\simeq Z(\Phi) r_c(\Phi)  F_c(\Phi)$ in the zero shear limit,
where $Z(\Phi)$ is the average coordination number,
$r_c(\Phi)$ and $F_c(\Phi)$ are respectively the average distance between contacting grains and the average force acting on 
the contact point.
 It is obvious that $Z(\Phi)$ and $r_c(\Phi)$ can be replaced by $Z(0)$ and $r_c(0)=\sigma$ 
in the vicinity of the jamming point, where $\sigma$ is the average diameter
of the particles.
Indeed, O'Hern et al. \cite{OHern03} verified $Z(\Phi)-Z(0)\propto \Phi^{1/2}$ for three dimensional cases.
Thus, the most important term is the mean contact force $F_c(\Phi)\propto \delta(\Phi)^{\Delta}$, where $\delta(\Phi)$ is the average
length of compression. 
Now, let us compress  the system at the critical point $\phi_J$ into $\Phi = \phi-\phi_J >0$ by an affine transformation. 
Since all the characteristic lengths are scaled by the system size, we may assume the approximate relation  
$r_c(\Phi) = (\phi_J/\phi)^{1/D}  \sigma$.
From the relation $\delta(\Phi) =  r_c(0)-r_c(\Phi)$,
 $\delta(\Phi)$ approximately satisfies 
$\delta(\Phi) \simeq (\sigma/D\phi_J)\Phi\sim \Phi$
 in the vicinity of $\Phi = 0$.
Thus, we conclude $P\sim \Phi^{\Delta}$.
This relation has also been verified  in Ref.~\citen{OHern03}. \

On the other hand, it is well-known that there is Coulomb's frictional law in granular systems in which
$S/P$ is a constant. Indeed, Hatano \cite{hatano07}  simulated
the sheared granular system under a constant pressure $P$ and demonstrated that 
the ratio satisfies 
$\lim_{\dot{\gamma}\rightarrow 0} S(\dot{\gamma},P)/P = S_Y(P)/P= M_0$, where 
$ S_Y(P) \equiv \lim_{\dot{\gamma}\rightarrow 0} S(\dot{\gamma},P) $
and 
the constant  $M_0$ is independent of the pressure $P$.
The excess volume fraction $\Phi(\dot{\gamma},P)$ in this system is
a function of $\dot{\gamma}$ and $P$, but we also can express  the pressure
 as $P(\dot{\gamma},\Phi)$.
Since
  $M_0=\lim_{\dot{\gamma}\rightarrow 0} S(\dot{\gamma},P(\dot{\gamma},\Phi))/P(\dot{\gamma},\Phi)= S_Y(P(0,\Phi))/P(0,\Phi)$
is independent of $P(0,\Phi)$,
$M_0$ should be independent of $\Phi$.
Thus, we can conclude that $\Phi$ dependence of $S$
is the same as that of $P$ in the limit $\dot\gamma\to 0$. From this result,
 we obtain $y_{\Phi}=\Delta$ and
\begin{equation}\label{y_p}
y_{\Phi}=1 \quad {\rm for}{~}{\rm linear{~}spring{~}model}.
\end{equation} 
This result  also implies $y_{\Phi}={y'}_{\Phi}$ which is consistent with the numerical observation.\cite{Hatano08_1} \

The next relation is related to the density of state. Wyart et al. \cite{Wyart} demonstrated the followings
for unsheared assemblies of elastic soft spheres.
(i) The jamming is related to the appearance of the soft modes in the density of state. (ii) There is a plateau in 
the density of state in the vicinity of jamming transition. (iii) The cutoff frequency $\omega^*$ of the plateau is proportional
to $\sqrt{P}$.
From the argument in the previous paragraph, the applied pressure satisfies the relation $P\propto \Phi^{\Delta}$.
When we assume that the characteristic frequency $\omega$ in the limit
$\dot{\gamma} \rightarrow 0$ can be scaled 
by the cutoff frequency $\omega^*$, we may conclude
$\omega\sim |\Phi|^{1/2}$ for the linear spring model. Thus, we obtain $z_{\Phi}=\Delta/2$ or
\begin{equation}\label{z_p}
z_{\Phi}=1/2 \quad {\rm for}{~}{\rm linear{~}spring{~}model}.
\end{equation}

Finally, we consider the characteristic frequency $\omega$ in the unjammed phase
($\Phi<0$).
 In this phase, the characteristic frequency $\omega$ is estimated as $\omega 
 \sim 
\sqrt{T/m}/l(\Phi)$, where $l(\Phi)$ is the mean free path.
Note that $l(\Phi)$ may be evaluated as 
$(\sigma /D\phi_J) |\Phi|$ in the vicinity of the point J,
using the parallel argument to  $\delta(\Phi)$ for $\Phi>0$.
From the scalings in Bagnold's regime (\ref{bagnold}), we obtain 
$\omega\sim |\Phi|^{z_{\Phi}(1-z_{\gamma}^{-1})}\dot\gamma$ and 
$T\sim |\Phi|^{x_{\Phi}(1-2x_{\gamma}^{-1})}\dot\gamma^2$. Substituting these relations to $\omega\sim \sqrt{T/m}/l(\Phi)$,
 we obtain
\begin{equation}\label{bag}
z_{\Phi}(1-z_{\gamma}^{-1})-\frac{1}{2}x_{\Phi}(1-2x_{\gamma}^{-1})=-1.
\end{equation}


From the above six relations \eqref{z_phi}-\eqref{bag} we finally determine the six critical exponents
\begin{eqnarray}\label{exponents}
x_\Phi& = &3, \quad x_\gamma=\frac{6}{5},
\quad y_\Phi=1, \nonumber \\ 
 y_\gamma& = & \frac{2}{5},
\quad z_\Phi=\frac{1}{2}, \quad z_\gamma=\frac{1}{5}
\end{eqnarray}
for the linear spring model. The exponents in Hertzian model are, of course, different. In general situation
for $\Delta$, eqs. (\ref{exponents}) are replaced by
\begin{eqnarray}\label{g-exponents}
x_{\Phi}&=&2+\Delta, \quad x_{\gamma}=\frac{2\Delta+4}{\Delta+4}, \quad y_{\Phi}=\Delta,
\nonumber\\
 y_{\gamma}&=&\frac{2\Delta}{\Delta+4}, \quad
z_{\Phi}=\frac{\Delta}{2}, \quad 
z_{\gamma}=\frac{\Delta}{\Delta+4}.
\end{eqnarray} 
We should note that the exponents are independent of the spatial dimension. This is not surprising because 
our phenomenology to derive eqs. (\ref{y_p}) and (\ref{z_p})
is independent of the spatial dimension. \cite{OHern03, Wyart}
We should stress an interesting feature of jamming transition that the exponents strongly depend on the interaction model among particles. This property is contrast to
 that in the conventional critical phenomena. 
Thus, we should be careful to use the idea of the universality in describing the jamming transition. 

From now on, let us verify our theoretical results based on the molecular dynamics simulation.
In our simulation, the system 
consists of $N$  spherical grains in $2,3,4$ dimensions.
We adopt the linear spring model ( $\Delta=1$ ) for simplicity. 
We also introduce dissipative force $-\eta \delta v$, where $\delta v$ 
represents the relative velocity 
between the contacting  particles. 
Each grain has an identical mass $m$.
In order to realize an uniform  
velocity gradient $\dot\gamma$ in  $y$ direction and macroscopic velocity only 
in the $x$ direction, we adopt
the Lees-Edwards boundary conditions. The particle 
diameters are $0.7\sigma_0$, $0.8\sigma_0$, $0.9\sigma_0$ and $\sigma_0$ each of which is
assigned to $N/4$ particles.

In our simulation $m$, $\sigma_0$ and $\eta$ are set to be unity, 
and all quantities are converted to dimensionless forms,
where the unit of time scale is $m / \eta$.
We use the spring constant $k = 1.0$.
For the system near the critical density, such as $\phi=0.8428$ for $D=2$,
$\phi=0.643$ and $0.6443$ for $D=3$, we use $N=4000$ in order to remove 
finite size effects.
For other systems, we use $N=2000$.

\begin{figure}
\begin{center}
\includegraphics[height=45em]{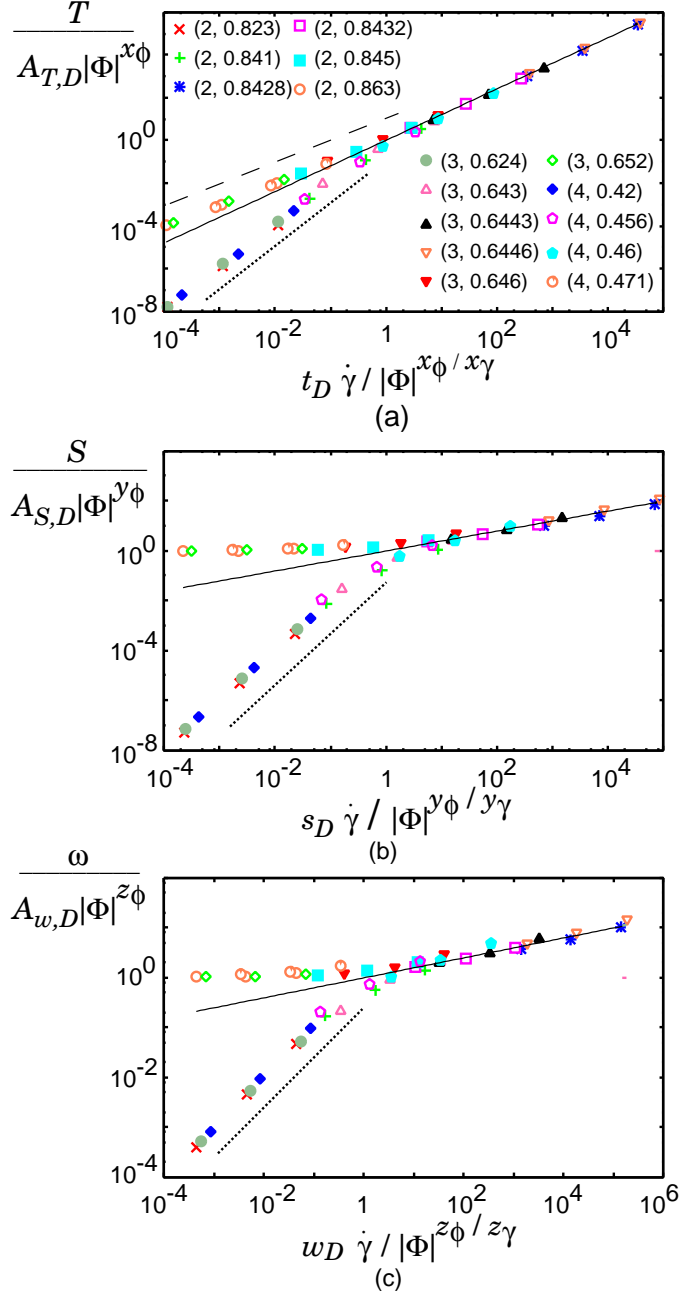}
\caption{ (Color online)
(a): Collapsed data of the shear rate dependence of the granular temperature $T$
using the scaling law for $D=2,3$ and $4$.
The dashed line, the dotted line and the solid line are proportional to 
$\dot{\gamma}$, $\dot{\gamma}^2$ and $\dot{\gamma}^{x_\gamma}$.
 The legends show the dimension $D$ and the volume fraction $\phi$
 as $(D, \phi)$.
(b): Collapsed data of the shear rate dependence of the shear stress $S$
using the scaling law for $D=2,3$ and $4$.
The dotted line and the solid line are proportional to 
 $\dot{\gamma}^2$ and $\dot{\gamma}^{y_\gamma}$.
(c): Collapsed data of the shear rate dependence of the cooling rate $\omega$
using the scaling law for $D=2,3$ and $4$.
The dotted line and the solid line are proportional to 
 $\dot{\gamma}^2$ and $\dot{\gamma}^{z_\gamma}$.
  }
\label{T_S_w:fig}
\end{center}
\end{figure}

The scaling plots of our simulation based on the exponents \eqref{exponents} are shown in Fig. \ref{T_S_w:fig}.
We should stress that these scaling plots contain the data in $D=2,3$ and 4.
The volume fraction at the point J is estimated as
$\phi_J = 0.84285$ for $D=2$, $\phi_J = 0.64455$ for $D=3$
or $\phi_J = 0.4615$ for $D=4$. 
We examine the shear rate $\dot{\gamma}$ is in the range between 
$5\times 10^{-7}$ and $5 \times 10^{-5}$ for $D=2,3$ and between $5\times 10^{-6}$ and $5 \times 10^{-4}$ for $D=4$.
The amplitudes and the adjustable parameters are obtained as
$(t_D,A_{t,D},s_D,A_{s,D},w_D,A_{w,D}) = (0.0125, 7.17, 0.025, 0.035, 0.05, 0.3)$ for $D=2$,   
$(0.01385, 2.527, 0.03, 0.04, 0.065, 0.65)$ for $D=3$,   
$(0.015, 1.6275, 0.03, 0.06, 0.06, 1)$ for $D=4$.   
Since Fig.\ref{T_S_w:fig} exhibits beautiful scaling laws,
 our phenomenology seems to be right. 

Figure 2(a) shows $\Phi$ dependence of the shear viscosity $\mu \equiv S/\dot{\gamma}  $ in 
Bagnold's regime. 
We should note that there is no consensus in previous studies on the shear viscosity. 
For example, Garcia-Rojo et al. \cite{Garcia} reported $\mu\sim 1/(\phi_c-\phi)$,
where $\phi_c$ is lower than $\phi_J$, while Losert et al. \cite{Losert}
 observed the exponent larger than 1 from their experiment, and the exponent of divergence in Ref.~\citen{Olsson} is also
between 1 and 2.
We also note that the viscosity is believed to diverge as $|\Phi|^{-2}$ for colloidal suspensions.\cite{Russel} \
However, our scaling theory predicts $\mu \propto 
|\Phi|^{y_\phi(1 - 2/y_\gamma)} \propto |\Phi|^{-4}$, 
and the viscosity diverges at the point J. 
Here, the scaling exponents for $\mu$ is independent of $\Delta$
because $y_\phi$ and $y_\gamma$ 
for an arbitrary $\Delta$ are determined from our theory as in eq. (\ref{g-exponents}).
As we can see in Fig. 2(a), the theoretical prediction is consistent with our numerical result.
We also examine the possibility that the viscosity diverges at $\phi_c<\phi_J$ with $\mu\sim (\phi_c-\phi)^{-1}$ as
in the case of Garcia-Rojo et al.\cite{Garcia} Actually we can fit the data of our two-dimensional simulation 
by $\mu\sim (\phi_c-\phi)^{-1}$ with $\phi_c=0.835$ which is less than $\phi_J=0.8428$ for $\phi<\phi_c$,
 but the viscosity is still finite even for $\phi>\phi_c$ (see Fig. 2(b)). 
Thus, we can conclude that (i) the viscosity does not satisfy $(\phi_c-\phi)^{-1}$ but exhibits a consistent
behavior with $(\phi_J-\phi)^{-4}$ predicted by our phenomenology, and
(ii) the critical behaviors are  only characterized by the point J.

We also verify the validity of $\omega\sim |\Phi|^{1/2}$ in the jammed phase 
from our simulation in Fig. 3. 
The envelope line of our result seems to be consistent
with the theoretical prediction.

It should be noted that the existence of the plateau for $|\Phi| \rightarrow 0$
in Fig. 2(a) can be understood from the scaling relation (\ref{S:scale}).
Indeed, $\Phi$ dependence of the shear stress 
disappears in the limit of large $x\equiv \dot \gamma /|\Phi|^{y_{\Phi}/y_{\gamma}}$ with 
${\cal S}_{-}(x)\to x^{y_{\gamma}}$ as in eq.(\ref{critical}). Thus, we obtain
$\mu/\dot\gamma=S/\dot\gamma^2\sim \dot\gamma^{y_{\gamma}-2}\sim \dot\gamma^{-8/5}$.
This estimation might be consistent with the simulation 
in Fig. 2(a) in which the value of the plateau increases
as the shear rate decreases. 
Similarly, we can expect the value of plateau of $\omega$ as $\dot\gamma^{z_{\gamma}}$,
 while this saturation 
cannot be verified from the simulation. 

\begin{figure}
\begin{center}
\includegraphics[height=14em]{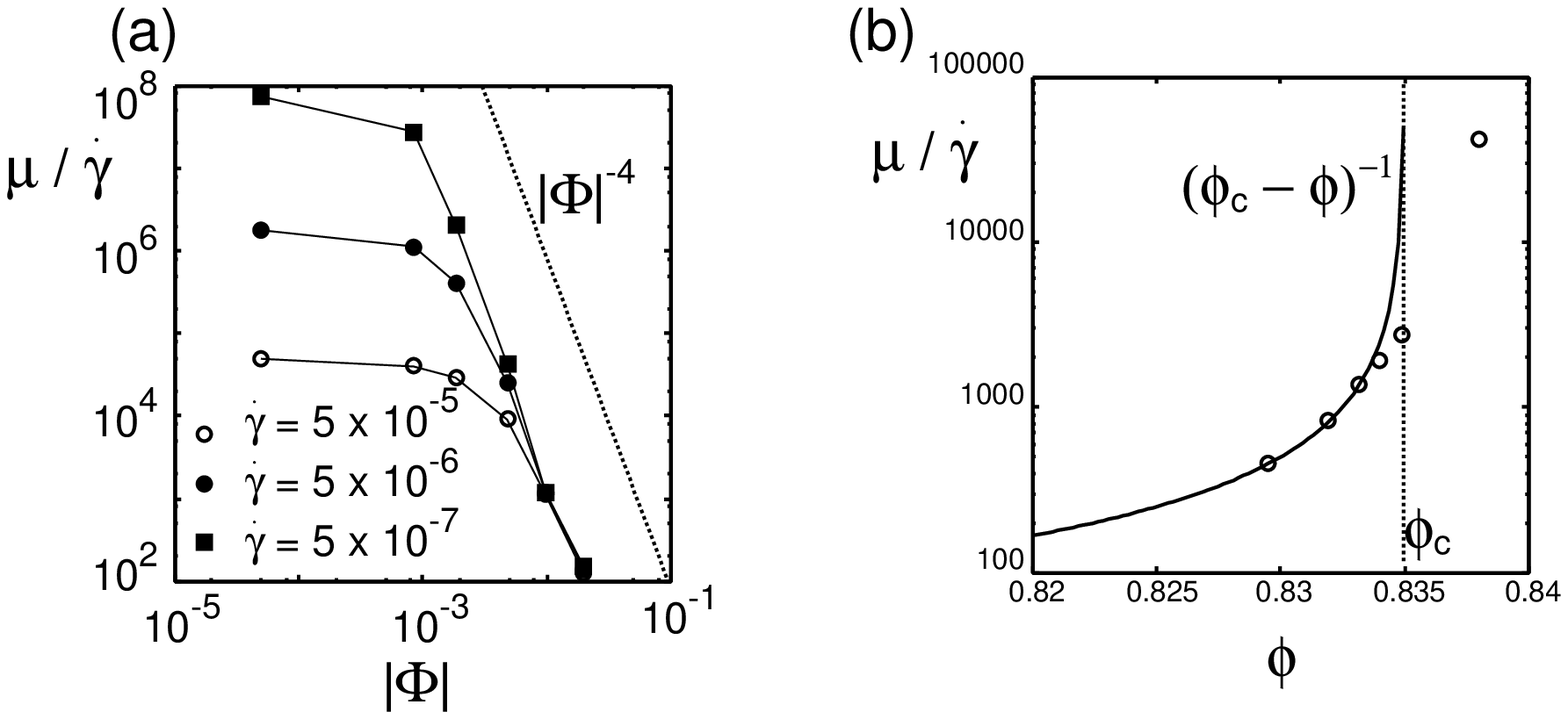}
\caption{ (a) :
$\mu/ \dot{\gamma}$ as a function of $\Phi$ for $D=2$
with $\dot{\gamma}=5\times 10^{-5}, 5\times 10^{-6}, 5\times 10^{-7}$
in the unjammed phase.
(b) :  $\mu/ \dot{\gamma}$ as a function of
$\phi$ for $\dot \gamma = 5 \times 10^{-7}$
in the unjammed phase,
 where the solid line is proportional to $(\phi_c-\phi)^{-1}$
 with $\phi_c=0.835$.
 }
\label{eta:fig}
\end{center}
\end{figure}

\begin{figure}
\begin{center}
\includegraphics[height=14em]{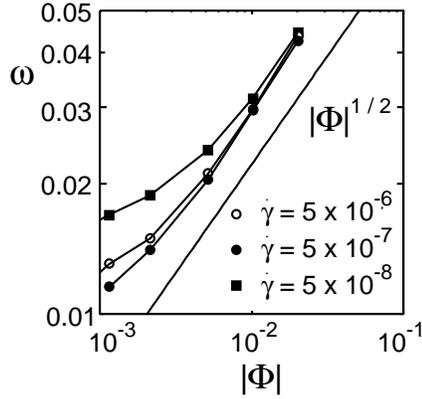}
\caption{ 
 $\omega$ as a function of $\Phi$ for $D=2$
with $\dot{\gamma}=5\times 10^{-6}, 5\times 10^{-7}, 5\times 10^{-8}$
in the jammed phase.
  }
\label{omega:fig}
\end{center}
\end{figure}

Now, let us discuss our results.
First of all, 
the ratios between the exponents $x_\phi / x_\gamma$, $y_\phi / y_\gamma$,
and $z_\phi / z_\gamma$ obtained in eq. (\ref{exponents}) or (\ref{g-exponents}) satisfy
\begin{equation}
\alpha\equiv \frac{x_\phi}{  x_\gamma} = \frac{y_\phi} { y_\gamma}
 = \frac{z_\phi} { z_\gamma} = \frac{\Delta +4}{2}.
\label{rate:eq}
\end{equation}
This is not surprising because the time scale is expected to be scaled by the shear rate. Thus,
the ratio $\alpha$ in eqs. (\ref{T:scale}),
(\ref{S:scale}), and (\ref{w:scale}) should be common.
In other words, the characteristic time scale $\tau$
exhibits the critical slowing down as $\tau \sim |\Phi|^{-\alpha}$.
This property has already been indicated by Hatano.\cite{Hatano08_1}.
Once we accept the ansatz  (\ref{rate:eq}),
eqs. (\ref{z_phi}),  (\ref{zyx}),  and (\ref{z_g}) are degenerate, and  reduce to 
\begin{equation}
x_\phi-y_\phi+z_\phi=\alpha.
\end{equation}
Equation (\ref{bag}) is also reduced to the simplified form
\begin{equation}
z_\phi = \frac{x_\phi}{ 2} -1.
\label{bag2}
\end{equation}
From these equations and eqs. (\ref{y_p}) and (\ref{z_p})
 with (\ref{rate:eq}),
we 
obtain eq. (\ref{exponents}) or eq. (\ref{g-exponents}).

Second, Hatano estimated the exponents 
$x_{\Phi}=2.5$, $x_{\gamma}=1.3$, $y_{\Phi}=1.2$ and $y_{\gamma}=0.57$ from his
three-dimensional simulation for the linear spring model\cite{Hatano08_1}, 
which differ from
our prediction (\ref{exponents}). 
In particular, if we use these values with eq. (\ref{bag2}),
 $z_\phi$ is estimated as $z_\phi= 0.25$, which is one half of our prediction $z_{\Phi}=1/2$. 
 However, the estimation of the scaling exponents
strongly depend on the choice of $\phi_J$ and the range of the shear rate $\dot{\gamma}$.
The value of $\phi_J$ and the range of $\dot{\gamma}$ in Ref.~\citen{Hatano08_1}
are larger than ours. 
If we adopt Hatano's $\phi_J$ and the range of $\dot{\gamma}$, 
our numerical data can be scaled by Hatano's scaling. Although his scaling can be used in the wide range of $\dot\gamma$,
 the deviation from his scaling can be detected in the small $\dot\gamma$  region
 ($\dot\gamma < 10^{-4}$).  It is obvious that 
we should use smaller $\dot\gamma$ as possible as we can to extract the critical properties.
This suggests that our exponents are more appropriate than Hatano's exponents in characterizing the jamming transition.
 The difficulty in determination of the exponents from the simulation 
also supports the significance of our theory to determine the scaling laws.

Third, the exponents obviously depend on the model of interaction between particles
as predicted in eq. (\ref{g-exponents}). 
Our preliminary simulation suggests that 
the numerical exponents for Hertzian contact model is consistent with the prediction of (\ref{g-exponents}). 
The numerical results on $\Delta$ dependence of the exponents will be reported
elsewhere.

Fourth, our results should be modified when we analyze the model in the zero temperature limit of Langevin thermostat.
This situation corresponds to that in Ref.~\citen{Olsson}. 
In this case, we should replace Bagnold's law in unjammed 
phase by Newtonian law $S\propto \dot\gamma$. 
As a result, all the scaling exponents have different values. We will discuss
the results of this situation elsewhere.

Finally, 
we comment on the relation between our results and the previous studies on dynamical heterogeneity in glassy materials.
The dynamical heterogeneity in glassy materials is characterized by the large fluctuations of four point correlation
function, in which the result strongly depends on the spatial dimension. 
On the other hand, our theory and numerical
simulation suggest that the critical fluctuation is not important and our phenomenology works well. 
Since the quantities we analyzed in this letter are
not directly related to the four-point correlation functions,
there is no distinct contradiction between them. 
To study the roles of critical fluctuations and dynamical heterogeneity 
we may need a more sophisticated theory. This will be our future task.
  
In conclusion, we develop the phenomenological theory in describing the jamming transition. We determine the critical
exponents which are independent of the spatial dimension. The validity of our theory has been verified by the molecular
dynamics simulation.

\section*{Acknowledgements}
We thank T. Hatano, H. Yoshino and S. Sasa for the valuable discussion.
This work is partially supported by Ministry of Education, Culture, Science and Technology (MEXT), Japan (Grant No. 18540371) and the Grant-in-Aid for the global COE program
"The Next Generation of Physics, Spun from Universality and Emergence"
from the Ministry of Education, Culture, Sports, Science and
Technology (MEXT) of Japan.
One of the authors (M. O.) thanks the Yukawa Foundation for the financial
support.
The numerical calculations were carried out on Altix3700 BX2 at YITP in Kyoto University.

%

\end{document}